\documentclass[12pt]{article}
\newcommand{\be}{\begin{eqnarray}}
\newcommand{\ee}{\end{eqnarray}}

\newcommand{\es}[2]{e^{\,\,\,#1}_{(10)#2}}
\newcommand{\Es}[2]{E^{\,\,\,#1}_{(11)#2}}
\newcommand{\psiten}[1]{\psi^{(10)}_{#1}}
\newcommand{\psitenu}[1]{\psi^{(10)#1}}
\newcommand{\psielf}[1]{\psi^{(11)}_{#1}}
\newcommand{\psielfu}[1]{\psi^{(11)#1}}
\newcommand{\psicos}[1]{\psi_{#1}}
\newcommand{\psicosu}[1]{\psi^{#1}}
\newcommand{\Psicos}{\Psi}
\newcommand{\oelf}[1]{\omega^{(11)}_{#1}}
\newcommand{\felf}[1]{F^{(11)}_{#1}}
\newcommand{\cDvs}{\stackrel{{\rm vs}}{{\cal{D}}}\!}
\newcommand{\cQvs}{\stackrel{{\rm vs}}{{\cal{Q}}}\!}
\newcommand{\Rvs}[1]{{\stackrel{{\rm{vs}}}{J}{}^{\!#1}}}
\newcommand{\Rs}[1]{\stackrel{{\rm s}}{J}{}^{\!#1}}

\newcommand{\Rsb}[1]{\stackrel{{\rm s}}{\bar{J}}{}^{\!#1}}
\newcommand{\Jvs}[2]{\stackrel{{\rm vs}}{J}{}^{\!(#1)}_{\!\!#2}}
\newcommand{\Otten}[1]{\tilde{\Omega}^{(10)}_{#1}}
\newcommand{\Oten}[1]{\Omega^{(10)}_{#1}}

\usepackage{amsfonts,amsmath}
\usepackage{latexsym}

\def\half{\frac12}
\def\G{\Gamma}

\def\p{\partial}
\def\d{\delta}
\def\L{\Lambda}

\def\o{\omega}
\def\th{\theta}
\def\S{\Sigma}

\def\nn{\nonumber}
\def\eps{\epsilon}

\def\cD{{\cal{D}}}
\def\cE{{\cal{E}}}
\def\cF{{\cal{F}}}
\def\cQ{{\cal{Q}}}
\def\cP{{\cal{P}}}
\def\cV{{\cal{V}}}
\def\Jh{\hat{J}}
\def\KE{K(E_{10})}

\def\lae{{\mathfrak{e}}}

\def\lak{{\mathfrak{k}}}
\def\llra{\leftrightarrow}

\begin{document}
{\flushright IHES/P/05/53\\AEI-2005-187\\[10mm]}

\begin{center}
{\bf \Large Hidden symmetries and the fermionic sector of
  eleven-dimensional supergravity}\\[7mm]
Thibault Damour\footnotemark[1], Axel Kleinschmidt\footnotemark[2] and
  Hermann Nicolai\footnotemark[2]\\[5mm]
\footnotemark[1]{\sl  Institut des Hautes Etudes Scientifiques\\
     35, Route de Chartres, F-91440 Bures-sur-Yvette, France}\\[3mm]
\footnotemark[2]{\sl  Max-Planck-Institut f\"ur Gravitationsphysik\\
     Albert-Einstein-Institut \\
     M\"uhlenberg 1, D-14476 Potsdam, Germany} \\[7mm]
\begin{minipage}{12cm}\footnotesize
\textbf{Abstract:} We study the hidden symmetries of the
  fermionic sector of $D=11$ supergravity, and the r\^ole of
  $K(E_{10})$ as a generalised `R-symmetry'. We find a consistent model
  of a massless spinning particle on an $E_{10}/\KE$ coset manifold whose
  dynamics can be mapped onto the fermionic and bosonic dynamics of
  $D=11$ supergravity in the near space-like singularity limit. This
  $E_{10}$-invariant superparticle dynamics might provide the basis of
  a new definition of M-theory, and might describe the `de-emergence'
  of space-time near a cosmological singularity.\\
\end{minipage}
\end{center}

Eleven-dimensional supergravity (SUGRA${}_{11}$) \cite{CJS} is believed
to be the low-energy limit of the elusive `M-theory', which is,
hopefully, a unified framework encompassing the various known string
theories. Understanding the symmetries of SUGRA${}_{11}$ is therefore
important for reaching a satisfactory formulation of M-theory. Many
years ago it was found that the toroidal dimensional reduction of
SUGRA${}_{11}$ to lower dimensions leads to the emergence of
unexpected (`hidden') symmetry groups, notably $E_7$ in the reduction
to four non-compactified spacetime dimensions \cite{CJ79}, $E_8$
in the reduction to $D=3$ \cite{CJ79,Ju83,MaSchw83,Ni87a}, and the affine
Kac--Moody group $E_9$ in the reduction to $D=2$ \cite{Ju81,Ni87b}. It was
also conjectured \cite{Ju85} that the hyperbolic Kac--Moody group
$E_{10}$ might appear when reducing SUGRA${}_{11}$ to only {\em one}
(time-like) dimension.

Recently, the consideration, {\em \`a la} Belinskii, Khalatnikov and
Lifshitz \cite{BKL}, of the {\em near space-like singularity
limit}\footnote{This limit can also be viewed as a small tension
  limit, $\alpha'\to\infty$.} of generic {\em inhomogeneous} bosonic
eleven-dimensional supergravity solutions has uncovered some striking
evidence for the hidden r\^ole of $E_{10}$
\cite{DaHe01,DaHeNi02}. Ref. \cite{DaHeNi02} related the gradient
expansion ($\p_x\ll\p_t$), which organises the near space-like
singularity limit \cite{DaHeNi03}, to an algebraic expansion in the
height of positive roots of $E_{10}$. A main conjecture of \cite{DaHeNi02}
was the existence of a {\em correspondence} between the time evolution,  
around any given spatial point ${\bf x}$, of the supergravity bosonic fields
$g_{MN}^{(11)}(t,{\bf x})$, $A_{MNP}^{(11)}(t,{\bf x})$, together with
their infinite towers of spatial gradients, on the one hand, and the
dynamics of a structureless massless particle on the
infinite-dimensional coset space $E_{10}/\KE$ on the other hand.
Here, $\KE$ is the maximal compact subgroup of $E_{10}$. Further evidence 
for the r\^ole of the
one-dimensional non-linear sigma model $E_{10}/\KE$ in M-theory was
provided in \cite{KlNi04a,KlNi04b,DaNi04,DaNi05}. 

An earlier and 
conceptually different proposal aiming at capturing hidden symmetries of 
M-theory, and based on the very-extended Kac--Moody group $E_{11}$,
was made in \cite{We00,We01} and further developed in
\cite{SchnWe01,SchnWe02,We03}. A proposal combining the
ideas of \cite{We01} and \cite{DaHeNi02} was put forward in
\cite{EnHo04a,EnHo04b,EnHeHo05}. 

In this letter, we extend the bosonic coset construction of \cite{DaHeNi02} 
to the full supergravity theory by including fermionic variables; more
specifically, we provide evidence for the existence of a correspondence 
between the time evolution of the coupled supergravity fields
$g_{MN}^{(11)}(t,{\bf x})$, $A_{MNP}^{(11)}(t,{\bf x}), \psielf{M}(t,{\bf
  x})$ and the dynamics of a {\em spinning massless particle} on
$E_{10}/\KE$. Previous work on $E_{10}$ which included fermions can be
found in \cite{KlNi04a,dBHP05}.\footnote{Results similar to some of the
 ones reported here have been obtained in \cite{dBHPin}.}

To motivate our construction of a fermionic extension of the bosonic
one-dimensional $E_{10}/\KE$ coset model we consider the equation of
motion of the gravitino in $D=11$ supergravity
\cite{CJS}.\footnote{We use the mostly plus signature;
  $M,N,\ldots=0,\ldots,10$ denote spacetime coordinate (world)
  indices; $m,n,p,\ldots =1,\ldots,10$ denote spatial coordinate
  indices, and the indices $i,j,k,l =1,\ldots,10$ label the
  non-orthonormal frame components $\theta^i{}_m dx^m$.
   Spacetime Lorentz (flat) indices are denoted
  $A,B,C,\ldots, F =0,\ldots, 10$, while $a,b,\ldots,f =1,\ldots,10$
  denote purely spatial Lorentz indices. We use the conventions of
  \cite{CJS,CJ79} except for the replacement $\G^M_{\rm{CJS}}=+i
  \G^M_{\rm{here}}$ (linked to the mostly plus signature) which allows
  us to use real gamma matrices and real (Majorana) spinors. The
  definition of the Dirac conjugate is $\bar{\psi} := \psi^T \G^0_{\rm{here}}$,
  and thus differs from \cite{CJS} by a factor of $i$.
  The field strength $F^{\rm{here}}_{MNPQ}$ used in this letter is equal
  to $+1/2$ the one used in \cite{DaHeNi02}.} 
Projecting all coordinate indices on an elfbein
$E^A_{(11)}= \Es{A}{M} dx^M$, the equation of motion for
$\psielf{A}=\Es{M}{A} \psielf{M}$ is (neglecting quartic
fermion terms)
\be\label{rseom}
0= \hat{\cE}_A := \G^B\left[\left(D_A(\omega)+\cF_A\right)\psielf{B} -
  \left(D_B(\omega)+\cF_B\right)\psielf{A}\right],
\ee
where $D_A(\omega) = \Es{M}{A} D_M$ denotes the moving-frame
covariant derivative 
$
D_A(\omega)\psielf{B} =
\p_A\psielf{B}+\oelf{A\,BC}\psielfu{C} +\frac14
\oelf{A\,CD}\G^{CD}\psielf{B},
$
and where $\cF_A := +\frac1{144}(\G_A{}^{BCDE}-8\d^B_A\G^{CDE})\felf{BCDE}$ 
denotes the terms depending on the 4-form field strength
$\felf{MNPQ}=4\p_{[M}^{\ } A^{(11)}_{NPQ]}$. Here
$\oelf{A\,BC}=-\oelf{A\,CB}= \Es{M}{A}\oelf{M\,BC}$ denotes the moving 
frame components of the spin connection, with $\oelf{A\,BC} =
\frac12 (\Omega^{(11)}_{AB\,C} + \Omega^{(11)}_{CA\,B} -
  \Omega^{(11)}_{BC\,A})$, where  
$\Omega^{(11)}_{AB\,C}=-\Omega^{(11)}_{BA\,C}$ are the  coefficients
  of anholonomicity. 
Following \cite{DaHeNi02,DaNi04} we use a pseudo-Gaussian (zero-shift)
coordinate system $t,x^m$ and we accordingly decompose the elfbein
$E^A_{(11)}$ in separate time and space parts as $E^0_{(11)}= N dt,\,
E^a_{(11)}= \es{a}{m} dx^m$. We note that the zehnbein
$E^a_{(11)}=e_{(10)}^a$ is related to the non-orthogonal,
time-independent spatial frame $\th^{i}({\bf
  x})=\th^{i}{}_m({\bf x})dx^m$ used in \cite{DaHeNi02} via
$e_{(10)}^a=S^a{}_{i}\th^{i}$ \cite{DaNi04}.

Using the $D=11$ local supersymmetry to impose the relation
$
\psielf{0} =\G_0\G^a \psielf{a},
$
and defining $\cE_a:=N g^{1/4}\G^0\hat{\cE}_a$
(with $g^{1/2}=\det(\es{a}{m})$), we find that the spatial
components of the gravitino equation of motion (\ref{rseom}),
when expressed in terms of a rescaled  $\psiten{a}:=g^{1/4}\psielf{a}$,
take the following form
\be\label{RS1}
\cE_a &=& \partial_t\psiten{a} + \oelf{t\, ab} \psitenu{b}
         + \frac14 \oelf{t\, cd} \G^{cd} \psiten{a} \\
 &-& \frac1{12} \felf{tbcd} \G^{bcd} \psiten{a} \, - \,
  \frac23 \felf{tabc} \G^b \psitenu{c}\, + \, \frac16 \felf{tbcd}
  \G_{a}{}^{bc} \psitenu{d}\nn\\
&+& \frac{N}{144} \felf{bcde} \G^0 \G^{bcde} \psiten{a} \, + \,
\frac{N}9 \felf{abcd} \G^0 \G^{bcde} \psiten{e} \, - \,
\frac{N}{72} \felf{bcde} \G^0 \G_{abcdef} \psitenu{f} \nn\\
&+&  N(\oelf{a\,bc} - \oelf{b\,ac})\G^0\G^b  \psitenu{c}
   \, +\, \frac{N}2 \oelf{a\,bc} \G^0 \G^{bcd} \psiten{d}
   \, - \, \frac{N}4 \oelf{b\,cd} \G^0 \G^{bcd} \psiten{a}\nn\\
&+& N g^{1/4} \G^0 \G^b \Big( 2 \p^{\ }_a \psielf{b} - \p^{\ }_b \psielf{a}
             - \frac12 \oelf{c\, cb} \psielf{a} - \oelf{0\,0a} \psielf{b}
             + \frac12 \oelf{0\, 0b} \psielf{a} \Big).\nn
\ee
Refs. \cite{DaHeNi02,DaNi04} defined a {\em dictionary}
between the temporal-gauge bosonic supergravity fields $g^{(11)}_{mn}(t,{\bf
  x}), A^{(11)}_{mnp}(t,{\bf x})$ (and their first spatial gradients:
  spatial connection and magnetic 4-form) and the
four lowest levels
$h^i{}_a(t)$, $A_{ijk}(t)$, $A_{i_1\ldots i_6}(t)$,
$A_{i_0|i_1\ldots i_8}(t)$ of the infinite tower of coordinates
parametrising the coset manifold $E_{10}/\KE$. Here,  we extend this
dictionary to fermionic variables by showing that the rescaled, SUSY
gauge-fixed gravitino field $\psiten{a}$ can be identified with the
first rung of a
`vector-spinor-type' representation of $\KE$, whose Grassmann-valued
representation vector will be denoted by
$\Psicos=(\psicos{a},\psicos{\ldots}, \ldots)$.\footnote{By contrast,
  \cite{dBHP05} considered `Dirac-spinor-type' representations of
  $\KE$.} We envisage $\Psicos$ to be an infinite-dimensional
representation of $\KE$ which is decomposed into a tower of $SO(10)$
representations, starting with a vector-spinor one $\psicos{a}$.
Our labelling convention is that coset quantities, such as
$A_{ijk}$ or $\Psicos$ do not carry sub- or superscripts, whereas
supergravity quantities carry an explicit dimension label.

We shall give several pieces of evidence in favour of this identification
and of the consistency of this $\KE$ representation. As in the bosonic case, 
the correspondence $\psiten{a}(t,{\bf x}) \llra
\psicos{a}^{\rm{coset}}(t)\equiv\psicos{a}(t)$
is defined at a fixed, but arbitrary, spatial point ${\bf x}$. 
A dynamical system governing a `massless spinning particle' on 
$E_{10}/\KE$ will be presented as an extension of the coset dynamics
of \cite{DaHeNi02} and we will demonstrate the consistency of this
dynamical system with the supergravity model under this
correspondence. More precisely, we will first show how to 
consistently identify the Rarita--Schwinger equation (\ref{RS1}) with
a $\KE$-covariant equation 
\be\label{kers}
0 = \cDvs\Psicos(t) := \bigg(\p_t - \cQvs(t)\bigg)\Psicos(t).
\ee
This equation expresses the parallel propagation of the vector-spinor-type
`$\KE$ polarisation' $\Psicos(t)$ along the $E_{10}/\KE$ worldline of the
coset particle. Our notation here is as follows. A one-parameter
dependent generic group element of $E_{10}$ is denoted by
$\cV(t)$. The Lie algebra valued `velocity' of $\cV(t)$,
namely $v(t)=\p_t\cV \cV^{-1}\in\lae_{10}\equiv{\rm{Lie}}(E_{10})$
is decomposed into its `symmetric' and `antisymmetric' parts according to 
$
\cP(t):= v_{\rm{sym}}(t):=\frac12 (v(t)+v^T(t)),
\cQ(t):= v_{\rm{anti}}(t):=\frac12 (v(t)-v^T(t)),
$
where the transposition $(\cdot)^T$ is the generalised transpose of an 
$\lae_{10}$ Lie
algebra element $x^T:=-\o(x)$ defined by the Chevalley involution
$\o$ \cite{Ka90}. $\KE$ is defined as the set of `orthogonal elements'
$k^{-1}=k^T$. Its Lie algebra $\lak_{10}={\rm{Lie}}(\KE)$ is made of all
the antisymmetric elements of $\lae_{10}$, such as $\cQ$. 

The bosonic coset model of \cite{DaHeNi02} is invariant under a
global $E_{10}$ right action and a local $\KE$ left action
$\cV(t)\,\longrightarrow\,k(t)\cV(t)g_0$. Under the local $\KE$
action, $\cP$ varies covariantly as $\cP \longrightarrow k \cP
k^{-1}$, while $\cQ$ varies as a $\KE$ connection $\cQ \longrightarrow
k \cQ k^{-1} + \p_t k\,k^{-1}$, with $\p_t k\,k^{-1}\in\lak_{10}$
following from the orthogonality 
condition. The coset equation (\ref{kers}) will therefore be $\KE$
covariant if $\Psicos$ varies, under a local $\KE$ left action,
as a certain (`vector-spinor') linear representation
\be
\Psicos\,\longrightarrow\, \stackrel{{\rm{vs}}}{R}\!(k)\cdot \Psicos
\ee
and if $\cQvs\,$ in (\ref{kers}) is the value of $\cQ \in \lak_{10}$
in the same representation
$\stackrel{{\rm{vs}}}{R}$.
In order to determine the concrete form of $\cQvs$ in
the vector-spinor representation we need an explicit
parametrisation of the coset manifold $E_{10}/\KE$. 

Following \cite{DaHeNi02,DaNi04} we decompose the
$E_{10}$ group w.r.t. its $GL(10)$ subgroup. Then the
$\ell=0$ generators of $\lae_{10}$ are
$\mathfrak{gl}(10)$ generators $K^a{}_b$ satisfying the 
standard commutation relations 
$
\left[K^a{}_b, K^c{}_d\right] = \d^c_b K^a{}_d - \d^a_d K^c{}_b.
$
The $\lae_{10}$ generators at levels $\ell=1,2,3$ as $GL(10)$ tensors 
are, respectively, $E^{a_1a_2a_3} = E^{[a_1a_2a_3]}$, $E^{a_1\ldots
  a_6} = E^{[a_1\ldots a_6]}$, and  $E^{a_0|a_1\ldots a_8} =
E^{a_0|[a_1\ldots  a_8]}$, where the $\ell=3$ generator is also
subject to  $E^{[a_0|a_1\ldots a_8]}=0$.
In a suitable (Borel) gauge, a generic coset element $\cV\in
E_{10}/\KE$ can be written as $\cV = \exp(X_h)\,\exp(X_A)$ with
\be\label{V}
X_h &=& h^b{}_a K^a{}_b,\\
X_A &=& \frac1{3!}A_{a_1a_2a_3}E^{a_1a_2a_3} + \frac1{6!} A_{a_1\ldots
  a_6}E^{a_1\ldots a_6} +\frac1{9!}A_{a_0|a_1\ldots
  a_8}E^{a_0|a_1\ldots a_8}+\ldots.\nn
\ee
Defining $e^i{}_a:=(\exp h)^i{}_a=\d^i{}_a + h^i{}_a +\frac1{2!}
h^i{}_s h^s{}_a+\ldots$ and $\bar{e}^a{}_i:=(e^{-1})^a{}_i$ one finds that
the velocity $v\in\lae_{10}$ reads, expanded up to $\ell=3$,
\be\label{vex}
v &=& \bar{e}^b{}_i\,\p_t e^i{}_a K^a{}_b +
  \frac1{3!}e^{i_1}{}_{a_1}e^{i_2}{}_{a_2}e^{i_3}{}_{a_3} DA_{i_1i_2i_3}
    E^{a_1a_2a_3} \\
&&+ \frac1{6!}e^{i_1}{}_{a_1}\cdots e^{i_6}{}_{a_6}
    DA_{i_1\ldots i_6} E^{a_1\ldots a_6} 
+ \frac1{9!}e^{i_0}{}_{a_0}\cdots e^{i_8}{}_{a_8}
    DA_{i_0|i_1\ldots i_8} E^{a_0|a_1\ldots a_8}.\nn
\ee
Here, $DA_{i_1i_2i_3}=\p_t A_{i_1i_2i_3}$, and the more complicated
expressions for $DA_{i_1\ldots i_6}$ and $DA_{i_0|i_1\ldots i_8}$ were
given in \cite{DaHeNi02}. In the expansion (\ref{vex})
of $v$   one can think of the indices on the generators $K^a{}_b$
etc. as {\em flat} (Euclidean) indices. As for the indices on
$DA_{i_1i_2i_3}$ etc. the dictionary of \cite{DaHeNi02,DaNi04}
shows that they correspond to a time-independent
non-orthonormal frame $ \theta^i =\theta^i{}_m dx^m$.
The object $e^i{}_a=(\exp h)^i{}_a$ (which is the `square root' of the
contravariant `coset metric' $g^{ij}=\sum_a e^i{}_a e^j{}_a$) relates the
two types of indices, and corresponds to the inverse of the matrix
$S^a{}_i$ mentioned above.
The parametrization (\ref{V}) corresponds to a special choice
of coordinates on the coset manifold $E_{10}/\KE$.

We introduce the $\lak_{10}$ generators through
\begin{align}
J^{ab} &= K^a{}_b - K^b{}_a\,, &\quad J^{a_1a_2a_3}
  &= E^{a_1a_2a_3} - F_{a_1a_2a_3}\,,&\nn\\
J^{a_1\ldots a_6} &= E^{a_1\ldots a_6}- F_{a_1\ldots
  a_6}\,,&\quad
   J^{a_0|a_1\ldots a_8} &= E^{a_0|a_1\ldots a_8}-
  F_{a_0|a_1\ldots a_8}\,,&
\end{align}
where $F_{a_1a_2a_3} = (E^{a_1a_2a_3})^T$ etc., that is, with the
general normalization $J= E-F$. Henceforth, we shall
refer to $J^{ab},\,J^{a_1a_2a_3}$, $J^{a_1\ldots
  a_6}$, and $J^{a_0|a_1\ldots a_8}$ as being of `levels' $\ell=0,1,2,3$,
respectively. However, this `level' is not a grading of
$\lak_{10}$; rather one finds for commutators that $\left[\lak^{(\ell)},\,
  \lak^{(\ell')}\right] \subset \lak^{(\ell+\ell')} \oplus
\lak^{(|\ell-\ell'|)}$ (in fact, $\lak_{10}$ is neither a graded nor a
Kac--Moody algebra). Computing the 
antisymmetric piece $\cQ$ of the velocity $v$ we conclude that the
explicit form of the fermionic equation of motion (\ref{kers}) is
\be\label{RScos}
&&\left(\p_t- \half\bar{e}^b{}_i\p_t e^i{}_a \Rvs{ab} 
  - \half\cdot\frac1{3!}e^{i_1}{}_{a_1}\cdots
  e^{i_3}{}_{a_3}DA_{i_1\ldots i_3} \Rvs{a_1a_2a_3} \right.\nn\\
&&\quad\quad\left.-\half\cdot\frac{1}{6!}e^{i_1}{}_{a_1}\cdots
  e^{i_6}{}_{a_6}DA_{i_1\ldots i_6} \Rvs{a_1\ldots a_6}\right.\nn\\
&&\quad\quad\left.-\half\cdot\frac{1}{9!}e^{i_0}{}_{a_0}\cdots
  e^{i_8}{}_{a_8}DA_{i_0|i_1\ldots i_8} \Rvs{a_0|a_1\ldots
    a_8}+\ldots \right)\Psicos=0.
\ee
Here, $\Rvs{ab}:=\stackrel{\rm{vs}}{R}\!(J^{ab})$ etc. are the form
the $\lak_{10}$ generators take in the sought-for
 vector-spinor representation $\Psicos$. The crucial consistency
condition for $\Psicos$ to be a linear representation is that the generators
$\Rvs{ab}$ etc. (to be deduced below) should satisfy the abstract
$\lak_{10}$ commutation relations
\be\label{KE10}
\left[J^{ab}, J^{cd}\right] &=&
  \d^{bc}J^{ad}+\d^{ad}J^{bc}-\d^{ac}J^{bd}-\d^{bd}J^{ac}  \equiv
  4 \d^{bc}J^{ad}\nn\\ 
\left[J^{a_1a_2a_3}, J^{b_1b_2b_3}\right] &=& J^{a_1a_2a_3b_1b_2b_3} -
  18 \d^{a_1b_1}\d^{a_2b_2}J^{a_3b_3}\nn\\
\left[ J^{a_1a_2a_3}, J^{b_1\ldots b_6}\right] &=&
  J^{[a_1|a_2a_3]b_1\ldots
    b_6}- 5!\,\d^{a_1b_1}\d^{a_2b_2}\d^{a_3b_3}J^{b_4b_5b_6}\nn\\
\left[J^{a_1\ldots a_6}, J^{b_1\ldots b_6}\right] &=& -6\cdot
   6!\,\d^{a_1b_1}\cdots \d^{a_5b_5} J^{a_6b_6}+\ldots\nn\\
\left[J^{a_1a_2a_3}, J^{b_0|b_1\ldots b_8}\right] &=& -336\,\left(
  \d^{b_0b_1b_2}_{a_1a_2a_3}J^{b_3\ldots b_8} -
  \d^{b_1b_2b_3}_{a_1a_2a_3} J^{b_4\ldots b_8b_0} \right)+\ldots\nn\\
\left[J^{a_1\ldots a_6}, J^{b_0|b_1\ldots b_8}\right] &=&
  - 8!\,\left(\d^{b_0b_1\ldots b_5}_{a_1\ldots a_6} J^{b_6b_7b_8} -
  \d^{b_1\ldots b_6}_{a_1\ldots a_6} J^{b_7b_8b_0}\right)+\ldots\nn\\
\left[J^{a_0|a_1\ldots a_8}, J^{b_0|b_1\ldots b_8}\right]
  &=& -8\cdot 8!\,\left(\d^{a_1\ldots a_8}_{b_1\ldots b_8} J^{a_0b_0} -
  \d^{a_1\ldots a_8}_{b_0b_1\ldots b_7} J^{a_0b_8} - \d^{a_0a_1\ldots
  a_7}_{b_1\ldots b_8} J^{a_8b_0}\right.\nn\\
&&\quad \left.+ 8 \,\d^{a_0}_{b_0} \d^{a_1\ldots a_7}_{b_1\ldots b_7}
  J^{a_8b_8} +7 \d^{a_1}_{b_0} \d^{a_0a_2\ldots a_7}_{b_1\ldots b_7}
  J^{a_8b_8}\right)+\ldots
\ee
computed up to $\ell=3$ in the basis for $\lae_{10}$ used in
\cite{DaNi04}.  We use the flat Euclidean $\d^{ab}$ of $SO(10)$ to
raise and lower indices. As $SO(10)$ representation the generator
$J^{a_0|a_1\ldots a_8}$ is reducible with  irreducible components 
$\bar{J}$ and $\Jh$ defined by $\bar{J}^{a_1|a_2\ldots a_9} =
J^{a_1|a_2\ldots a_9} -\frac83 \d^{a_1[a_2} \Jh^{a_3\ldots a_9]}$
and $\Jh^{a_3\ldots a_9} = \d_{a_1a_2} J^{a_1|a_2a_3\ldots
  a_9}$. Neglecting $\bar{J}^{a_0|a_1\ldots a_8}$, the corresponding
commutators for $K(E_{11})$ were already computed 
in \cite{We03}.
In eq.~(\ref{KE10}) we have used a shorthand notation
where the terms on the r.h.s. should be antisymmetrised (with weight
one) according to the antisymmetries on the l.h.s., as written out for
the $SO(10)$ generators $J^{ab}$ in the first
line. For the mixed
symmetry generator $J^{a_0|a_1\ldots a_8}$ this includes only
antisymmetrisation over $[a_1\ldots a_8]$. 
Under $SO(10)$ the tensors on the higher levels
rotate in the standard fashion.

To compare eqs.~(\ref{RS1}) and (\ref{RScos}) we now use the bosonic
dictionary obtained in \cite{DaHeNi02,DaNi04}. In terms of our present
conventions, and in terms of `flat' indices on both
sides\footnote{ To convert `frame' indices $i,j,k,\ldots$ into `flat'
ones $a,b,c,\ldots$, one uses  $e^i{}_a$ on the coset side, and
$\es{i}{a} := \theta^i{}_m \es{m}{a} \equiv (S^{-1})^i{}_a$
on the SUGRA side}
this dictionary consists of asserting the correspondences
\begin{align}\label{corres}
e^i{}_a &\llra \theta^i{}_m \es{m}{a},& DA_{a_1a_2a_3}
&\llra 2 \felf{t a_1a_2a_3}= 2 N
   \felf{0 a_1a_2a_3},&\nn\\
DA_{a_1\ldots a_6}&\llra -\frac2{4!} N \eps_{a_1\ldots a_6b_1\ldots
  b_4}\felf{b_1\ldots b_4},& DA_{a_0|a_1\ldots a_8} &\llra
\frac32 N\eps_{a_1\ldots a_8b_1b_2}\Otten{b_1b_2\,a_0}.&
\end{align}
Here, as in \cite{DaNi04},
$\Otten{ab\,c}=\Oten{ab\,c} -\frac29\d_{c[a}^{\ }\Oten{b]d\,d}$
denotes the tracefree part of the spatial anholonomy coefficient
$\Oten{ab\,c}= 2 \es{m}{[a}\es{n}{b]}\p^{\ }_m\es{c}{n}$.

Using the correspondences (\ref{corres}), as well as their consequence
$-\frac12(\bar{e}^{b}{}_{i}\p_t e^{i}{}_{a}
- \bar{e}^a{}_i\p_t e^{i}{}_{b})
\llra + \oelf{t\,ab}=N\oelf{0\,ab}$, we can tentatively re-interpret most
terms in the supergravity equation (\ref{RS1}) as terms in the
putatively $\KE$ covariant equation (\ref{RScos}). Using, as is always
locally possible, a spatial frame such that the trace $\oelf{b\,bc}=0$
(and therefore $\Otten{ab\,c}=\Oten{ab\,c}$), and neglecting, as in the
bosonic case \cite{DaHeNi02}, the {\em
  frame} spatial derivatives 
$\p^{\ }_a\psiten{b}$ and $\p_a N=
-N\oelf{0\,0a}$, we can identify eq.~(\ref{RS1}) with
eq.~(\ref{RScos}) if we {\em define} the action of $\KE$ generators in
the vector-spinor representation by
\be\label{vstrm}
\left(\Jvs{0}{\L}\Psicos\right)_a &:=& \L_{ab}\psicosu{b}
   +\frac14\L_{bc}\G^{bc}\psicos{a} ,\nn\\
\left(\Jvs{1}{\L}\Psicos\right)_a &:=& \frac1{12}\L_{bcd} \G^{bcd} \psicos{a} +
  \frac23 \L_{abc} \G^b \psicosu{c} - \frac16 \L_{bcd} \G_{abc}
  \psicosu{d},\nn\\ 
\left(\Jvs{2}{\L}\Psicos\right)_a &:=& \frac1{1440} \L_{bcdefg}
  \G^{bcdefg} \psicos{a} + \frac1{180} \L^{bcdefg} \G_{abcdef}
  \psicos{g}\nn\\
&&\qquad  - \frac1{72} \L_{abcdef} \G^{bcde} \psicosu{f},\nn\\
\left(\Jvs{3}{\L}\Psicos\right)_a &:=&
   \frac23\cdot \frac1{8!} \bigg(\L_{b|c_1\dots c_8}
  \G_{a}{}^{c_1 \dots c_8} \psicosu{b}
     +8 \L_{a|c_1\dots c_8} \G^{c_1\dots c_7}\psicosu{c_8} \nn\\
&& +2 \L_{b|bc_1\dots c_7} \G^{c_1 \dots c_7} \psicos{a} -28
   \L_{b|bc_1\dots c_7} \G_a{}^{c_1 \dots c_6} \psicosu{c_7}\bigg)
\ee
Here, we have used a shorthand notation for the action of $\Rvs{}$
by absorbing the transformation parameters into the generators
according to 
$\Jvs{0}{\L}\equiv\frac12\L_{a_1a_2} \Rvs{a_1a_2}$, $\Jvs{1}\L\equiv
\frac1{3!}\L_{a_1a_2a_3}\Rvs{a_1a_2a_3}$, $\Jvs{2}\L\equiv
\frac1{6!}\L_{a_1\dots a_6}\Rvs{a_1\ldots a_6}$, and
$\Jvs{3}{\L}\equiv\frac1{9!}\L_{a_0|a_1\ldots a_8}\Rvs{a_0|a_1\ldots
  a_8}$. 
The last parameter $\L_{a_0|a_1\ldots a_8}$ has two
irreducible pieces analogous to $\Jvs{3}{}$ and the trace appears
explicitly in (\ref{vstrm}).  

Proving the $\KE$ covariance of the coset fermionic equation
(\ref{RScos}) now reduces to proving that the generators $\Jvs{\ell}{}$ 
defined by (\ref{vstrm}) do satisfy the $\KE$ commutation relations
which were given in (\ref{KE10}). It is easy to see that the
commutators of the
level-zero generators $\Jvs{0}{}$ with themselves, as well as with any
other $\Jvs{\ell}{}$ for $\ell > 0$, produce the required $SO(10)$ rotations of
(\ref{KE10}). The other commutators require some tedious calculations
using the gamma algebra. The result of this computation is
\be\label{ell1comm}
\left(\left[\Jvs{1}{\L},\Jvs{1}{\L'}\right]\Psicos\right)_a &=& 20
 \left( \Jvs{2}{\S}\Psicos\right)_a
 -\left(\Jvs{0}{\S}\Psicos\right)_a ,\nn\\
\left(\left[\Jvs{1}{\L},\Jvs{2}{\L'}\right]\Psicos\right)_a &=& 56
 \left( \Jvs{3}{\S}\Psicos\right)_a
 -\frac16\left(\Jvs{1}{\S}\Psicos\right)_a,
\ee
where the $\Jvs{\ell}{\S}$ are defined as above, but now with 
new parameters given by 
$\Sigma^{(0)}_{ab} = \L^{\ }_{d_1d_2[a}\L'_{b]}{}^{d_1d_2}$,
$\Sigma^{(2)}_{b_1\ldots b_6}=\L^{\ }_{[b_1b_2b_3} \L'_{b_4b_5b_6]}$,
$\Sigma^{(1)}_{a_1a_2a_3} = \L^{b_1b_2b_3}\L'_{b_1b_2b_3a_1a_2a_3}$, 
and $\Sigma^{(3)}_{a_0|a_1\ldots a_8} = \L^{\ }_{a_0[a_1a_2}
  \L'_{a_3\ldots a_8]} - \L^{\ }_{[a_1a_2a_3} \L'_{a_4\ldots
    a_8]a_0}$. 
One can now check that the relations (\ref{ell1comm}) are
consistent with the $\KE$ commutators (\ref{KE10}).
All other commutators have to produce terms on the r.h.s. which have
contributions of `level' $\ell>3$ and therefore cannot be checked
fully. However, we have verified, where possible,
that the expected contributions of
the lower levels appear with the correct normalisation required by the
structure constants of (\ref{KE10}). Therefore we find that the
vector-spinor representation $\Jvs{\ell}{}$ of $\KE$ which we deduced from
comparing (\ref{RS1}) and (\ref{RScos}) is a good linear representation 
up to the level we have supergravity data to define it.

Using arguments from the the general representation theory of Lie
algebras one can actually show that the checks we have carried out are
sufficient to guarantee the existence of an extension of the vector-spinor
representation $\Jvs{\ell}{}$ to `levels' $\ell>3$ on the same
components $\psicos{a}$. 
That is, we can define on $\psicos{a}$ alone an {\em
  unfaithful, irreducible} {\bf   320}-dimensional 
representation of $\KE$ on which infinitely many $\KE$ generators are
realised non-trivially. For this definition it is sufficient to define
the action of $\Jvs{0}{}$ and $\Jvs{1}{}$ on $\psicos{a}$ and check
Serre-type compatibility conditions \cite{Be89}. We view the fact that
the $\Jvs{2}{}$ 
and $\Jvs{3}{}$ transformations deduced from the supergravity
correspondence above agree with this general construction as strong
evidence for the relevance of the vector-spinor component of the
{\em infinite-dimensional} $\KE$ spinor $\Psicos=(\psicos{a},\ldots)$
we have in mind.
If one repeats the same analysis for the Dirac
spinor, where the representation matrices on this {\bf 32}-dimensional
space are given in terms of anti-symmetric $\G$-matrices (see
(\ref{drep}) below), one finds that one can consistently realise 
$\KE$ on a {\bf 32}-component spinor of $SO(10)$. The  fact that the 
anti-symmetric $\G$-matrices together with $\G^0$ span
the fundamental representation of $SO(32)$ has led a number of authors
to propose $SO(32)$ as a `generalised holonomy' for M-theory 
\cite{DuSt91,DuLiu03}. That this group, like the larger group SL(32)
proposed in \cite{Hu03}, cannot be realised as a {\it bona fide}
symmetry was subsequently  pointed out in \cite{Ke04} where it was shown
that no suitable {\em spinor} (i.e. double valued) representation with
the correct number of components  
of these generalised holonomy groups exist. Our approach is radically
different, since we have an action not of $SO(32)$ but of $\KE$, with
infinitely many generators acting in a non-trivial manner, on a {\em
  bona fide} spinor representation of $SO(10)$. We therefore evade the
conclusions of \cite{Ke04}.\footnote{The transition from $SO(10)$ to
  $SO(32)$ (or $SO(1,10)$ to $SL(32)$) requires $\G^{abc}$ which is
  associated with the rank three gauge field. The importance of the
  rank three generator in the context of M5-brane dynamics was already
  stressed in \cite{BaWe00} and also features in \cite{We03} where it
  is seen as part of $K(E_{11})$. However, it is an open question whether 
  there exists a vector-spinor-type representation of  $K(E_{11})$,
  which would be analogous to (\ref{vstrm}) and thus also compatible 
  with \cite{Ke04}.}
The appearance of an unfaithful representation for the
fermions was already noted and studied in the affine case for
$K(E_9)$, which shows very similar features consistent with our
present findings \cite{NiSa05}. One possibility to construct a faithful
representation of $\KE$ already pointed out there might be to
consider the tensor product of such unfaithful representations with a
faithful representation, like the adjoint $\lak_{10}$ or the coset
$\lae_{10}\ominus\lak_{10}$.\footnote{Let us also note that the {\bf
  320}-dimensional representation of $\KE$ is compatible with the 
fermionic representations studied in~\cite{KlNi04a}.}
More details on these aspects will be given in a future publication
\cite{DaKlNiin}. 

A deeper confirmation of the hidden $\KE$ symmetry of SUGRA${}_{11}$
is obtained by writing down a $\KE$ invariant action functional
describing a massless spinning particle on $E_{10}/\KE$. We will be
brief and defer the details to \cite{DaKlNiin}. The bosonic part of
the action is the one of \cite{DaHeNi02} 
\be
S_{\rm{bos}} = \int dt\, \frac1{4n}\langle \cP(t)|\cP(t)\rangle
\ee
where $\langle\cdot|\cdot\rangle$ is the standard invariant bilinear
form on $\lae_{10}$ \cite{Ka90} and where the coset `lapse' function $n$
can be identified with the rescaled supergravity lapse $N g^{-1/2}$
(denoted $\tilde{N}$ in \cite{DaHeNi03}).

The fermionic term we add to this action reads
\be
S_{\rm{ferm}} = - \frac{i}2 \int dt\, \left(\Psicos(t)|\cDvs
  \Psicos(t)\right)_{\rm{vs}},
\ee
where $(\cdot|\cdot)_{\rm{vs}}$ is a $\KE$ invariant symmetric
form on the vector-spinor
representation space. Observe that this symmetric
form is actually {\em anti}-symmetric when evaluated on {\em
anti}-commuting (Grassmann valued) fermionic variables $\Psicos
(t)$, such that e.g. $\left(\Psicos(t)| \Psicos(t)\right)_{\rm{vs}} =
0$. On the lowest component of $\Psicos=(\psicos{a},\ldots)$ it is
explicitly given by 
$
\left(\Psicos|\Phi\right)_{\rm{vs}}=\psicos{a}^T\G^{ab}\phi_b.
$
The invariance of this form under the generators $\Jvs{\ell}{}$
defined in (\ref{vstrm}) is a quite restrictive condition. We have
verified that invariance holds, but only since we are working over
a {\em ten-dimensional} Clifford algebra.
By using induction arguments we find that 
$\left(\Psicos|\Phi\right)_{\rm{vs}}$ 
is invariant not only under (\ref{vstrm}) but under the
(unfaithful) extension to the {\em full} $\KE$ transformations
mentioned above.  We expect that the form 
$\left(\Psicos|\Phi\right)_{\rm{vs}}$ 
will extend
to an invariant symmetric form on a faithful representation
$\Psicos=(\psicos{a},\ldots)$.

Further important hints of a hidden $\KE$ symmetry come from
considering the local SUSY constraint ${\cal S}^{(11)}=0$ which is
proportional to the time component of the Rarita Schwinger equation
(\ref{rseom}). First, we find that, under the dictionary of
\cite{DaHeNi02,DaNi04},  ${\cal S}^{(11)}$ is mapped  into a $\KE$
covariant constraint of the form $\cP\odot\Psicos =0$, when
neglecting frame gradients $\p_a\psicos{b}$ as we have done in the
derivation of (\ref{vstrm}). The product $\odot$ symbolises a map
from the tensor product of $\lae_{10}\ominus\lak_{10}$ with $\Psicos$
onto a Dirac-spinor-type representation space of
$\lak_{10}$. The coset constraint $\cP\odot\Psicos = 0$ suggests to
augment the action $S_{\rm{bos}}+S_{\rm{ferm}}$ by a `Noether' term of
the form  
\be
S_{\rm{Noether}} = \int dt\,
  \left(\chi(t)|\cP(t)\odot\Psicos(t)\right)_{\rm{s}},
\ee
with a local Dirac-spinor $\chi(t)$ Lagrange multiplier (that is,
a one-dimensional `gravitino'). The total action
$S_{\rm{bos}}+S_{\rm{ferm}}+S_{\rm{Noether}}$ is expected to be
not only invariant under $\KE$, but also (disregarding
$\Psi^4$ terms) under time-dependent supersymmetry transformations which
involve a 
Dirac-spinor-type $\KE$ representation $\eps(t)$. In this case the 
$\chi=0$ gauge fixed action will be invariant under residual {\em
  quasi-rigid} supersymmetry 
transformations constrained to
satisfy $\stackrel{{\rm{s}}}{\cD}\eps(t)\equiv(\p_t -
\stackrel{\rm{s}}{\cQ})\eps=0$. This equation is formally
the same as (\ref{kers}) and (\ref{RScos}) but now the generators are
found to be (cf. \cite{dBHP05})
\begin{align}\label{drep}
\Rs{ab} &= \frac12 \G^{ab},& \quad \Rs{a_1a_2a_3} &= \frac12
  \G^{a_1a_2a_3},&\nn\\
\Rs{a_1\ldots a_6} &= \frac12 \G^{a_1\ldots a_6},&\quad
  \Rs{a_0|a_1\ldots a_8} &= 12\, \d^{a_1\ldots
  a_8}_{a_0b_1\ldots b_7} \G^{b_1\ldots b_7}.&
\end{align}
The particular form of the Dirac-spinor representation on $\ell=3$
implies that the irreducible component $\Rsb{a_0|a_1\ldots a_8}$ is
mapped to zero under this correspondence: indeed, there is no way to 
represent a non-trivial Young tableau purely in  terms of gamma matrices.
This is in contrast to the vector-spinor representation (\ref{vstrm}).

In summary, we have given evidence for the following generalisation of
the correspondence conjectured in \cite{DaHeNi02}: The time evolution
of the eleven-dimensional supergravity fields $g_{MN}^{(11)}(t,{\bf
  x}), A_{MNP}^{(11)}(t,{\bf x}), \psielf{M}(t,{\bf x})$ and their
spatial gradients (considered around any given spatial point ${\bf
  x}$, in temporal gauge and with fixed SUSY gauge) can be mapped
onto the dynamics of a (supersymmetric) {\em spinning massless
  particle} $(\cV(t),\Psicos(t))$ on $E_{10}/\KE$. The
$E_{10}$-invariant quantum dynamics of this superparticle might
provide the basis of a new definition of M-theory. Much work remains
to be done to extend the evidence indicated here, for instance by
proving the existence of irreducible faithful (and hence
infinite-dimensional) `vector-spinor-type' and `Dirac-spinor-type'
representations of $\KE$. 

Let us finally note on the physical side, that we deem it probable that
the proposed correspondence between M-theory and the coset model is such 
that the two sides do not have a common range of physical validity: Indeed, 
the coset model description emerges in the near space-like singularity
limit $T\to 0$, where $T$ denotes the proper time\footnote{The
  coordinate and `coset time' $t$ used above is 
 (in the gauge $n=1$)  roughly proportional to $ - \log T$, and
  actually goes to $+\infty$ 
  near the space-like singularity.}, which indicates
that the coset description might be well defined only when $T\ll
T_{\rm{Planck}}$, i.e. in a strong curvature regime where the
spacetime description `de-emerges'.

{\bf Acknowledgements}\\
We thank Ofer Gabber and Pierre Vanhove for informative discussions
and Bernard Julia for clarifying the conventions of \cite{CJS,CJ79}.
AK and HN gratefully acknowledge the hospitality of IHES during
several visits. This work was partly supported by the European Research and 
Training Networks `Superstrings' (contract number MRTN-CT-2004-512194)
and `Forces Universe' (contract number MRTN-CT-2004-005104).

\end{document}